# AutOMP: An Automatic OpenMP Parallelization Generator for Variable-Oriented High-Performance Scientific Codes

Gal Oren[1,2], Yehuda Ganan[3,4], and Guy Malamud[2]

[1] Department of Computer Science, Ben-Gurion University of the Negev, P.O.B. 653, Be'er Sheva, Israel
[2] Department of Physics, Nuclear Research Center-Negev, P.O.B. 9001, Be'er-Sheva, Israel
[3] Department of Physics, Bar-Ilan University, IL52900, Ramat-Gan, Israel
[4] Israel Atomic Energy Commission P.O. B. 7061, Tel Aviv 61070, Israel

orenw@post.bgu.ac.il, yehuda.ganan@live.biu.ac.il, guy.malamud@gmail.com

**Abstract.** OpenMP is a cross-platform API that extends C, C++ and Fortran and provides shared-memory parallelism platform for those languages. The use of many cores and HPC technologies for scientific computing has been spread since the 1990's, and now takes part in many fields of research. The relative ease of implementing OpenMP, along with the development of multi-core shared memory processors (such as Intel Xeon Phi) makes OpenMP a favorable method for parallelization in the process of modernizing a legacy codes. Legacy scientific codes are usually holding large number of physical arrays which being used and updated by the code routines. In most of the cases the parallelization of such code focuses on loop parallelization. A key step in this parallelization is deciding which of the variables in the parallelized scope should be private (so each thread will hold a copy of them), and which variables should be shared across the threads. Other important step is finding which variables should be synchronized after the loop execution. In this work we present an automatic pre-processor that preforms these stages - AutOMP (*Automatic OpenMP*). AutOMP recognize all the variables assignments inside a loop. These variables will be private unless the assignment is of an array element which depend on the loop index variable. Afterwards, AutOMP finds the places where threads synchronization is needed, and which reduction operator is to be used. At last, the program provides the parallelization command to be used for parallelizing the loop.

## 1 Introduction

### 1.1 Computational Scientific Software Parallelization

In computational scientific computing, advanced computing capabilities are used in order to study and solve complex scientific problems. This field exists since the 1960's [1]. However, since parallel computing was introduced to the scientific community in the 1990's, it has been growing rapidly [2]. As of today, High Performance Computing (HPC) platforms are taking a major part in many fields research, such as physics, biology and chemistry. In fields such as fluid dynamics, a system of conservation equations (mass, momentum and energy) is solved, using a time and space discretization over the physical domain of the problem at hand [3]. In a naive parallel algorithm, the domain is divided to different computing nodes, each one responsible on solving a part of the domain. The deviation may use shared memory or disturbed memory [4].

As oppose to "text book" computer codes, scientific codes are usually "variable oriented", meaning that the code holds a large number of physical arrays, (e.g. velocity or internal energy), which are being shared, used and/or updated by the code routines [5]. For example, in a hydrodynamic code, each point in space (vertex) holds its Lagrangian coordinates and its Lagrangian velocity vector, along with scalars such as the hydrostatic pressure. In order to obtain the velocities for the next time step, the code has to solve the momentum equation (which include information from adjacent vertices), resulting with new, updated velocities. These velocities are then translated to a shift in the coordinates, influencing the internal energy and pressure [6]. The above-described orientation introduces a difficulty if ones try to implement a non-synchronized parallelization or try to run on a non shared-memory platform (such as NUMA architecture computers) [7].

## 1.2 OpenMP & Shared-Memory Architectures

OpenMP [8] is an arrangement of compiler directives and callable runtime library methods – but not a programming language – that augment Fortran, C and C++ to express shared-memory parallelism. Because of its nature as a library and not a programming language, it can be merged into a serial program to guide how the workload is to be partition between threads that will run on variant processors or cores, and to manage the access to the shared data as needed. Because OpenMP let the base language to not be defined, vendors can realize OpenMP in any Fortran compiler available. OpenMP is appropriate for application on a vast range of SMP architectures [9].

The accelerate necessity for compute power is emerging quickly in lots of research spectrums. Some well known accelerators, such as GPUs, are filling these requirements except of one: They usually demand an extensive rewrite of the application code using ad-hoc programming languages such as CUDA or OpenCL [10]. By contrast, the Intel Xeon Phi coprocessor is established on the Intel Many Integrated Core Architecture and can be programmed with common parallelization techniques such as OpenMP [11]. Therefore, a combination of OpenMP parallelism on Xeon Phi coprocessor is ideal for programs that cannot or should not be parallelized with MPI from one hand, and that need to be used on shared-memory platform on the other hand.

In the current work, we discuss parallelization using OpenMP, in order to accelerate and optimize the solution. The present work was done with the LEEOR2D code [12][13][14], developed to solve the Euler conservation system in a hydrodynamic continuum, and designated to run on a Xeon Phi coprocessors.

## 1.3 Background Factors Affecting Parallelization Efficiency in OpenMP

OpenMP library allows users to determine the parallelization method in each branch of their code. In this subsection, the factors affecting the efficiency of parallelization will be stated, as well as the principles the user should follow in order to optimally improve efficiency. The next subsection will demonstrate how these principles affect the parallelization commands of loops.

The basic challenge presented to parallel algorithms developers is the complete utilization of the computational infrastructure, which derives the challenge of correct and efficient load balancing. In other words, there is an advantage to a situation where the amount of computational operations performed by each one of the threads is similar, and such state should be reached. Clearly, the ultimate accomplishment in this sense is that the running time of each of the threads is (almost) equal. The OpenMP library provides a simple solution to this challenge using dynamic command processing. In this way, during the running time, the master thread sends tasks that are smaller in size than the total computation size to each one of the slave threads, and each time a slave thread finishes its previously sent task, it receives another task from the master thread, which additionally performs tasks by itself. The loads balancing in this method depends on the size of the smaller tasks compared to the whole task, and usually the balancing is relatively good. One of the main advantages of the OpenMP library compared to MPI library is the ease of achieving this state.

One of the problems that may be caused by dynamic parallelization is a situation where the amount of time needed for each thread to ask the master thread to send the task it needs to receive becomes considerably high compared to the total time of the whole task. This synchronization problem is called *granularity* and can be solved easily by increasing the size of the smaller task; this obtains *Coarse grain granularity*. However, this increase has a trade-off, and the quality of task assignment may be compromised. OpenMP allows task distribution methods that are more complex (such as loop parallelization in the *guided* command), which attempt to obtain a relatively balanced load balancing and as coarse grain as possible.

An additional aspect that the user should be aware of is to decrease the information synchronization as much as possible. At this point, it should be emphasized that since this is shared memory, writing at the same time by two threads may lead to undesirable results. OpenMP library allows the declaration of code sections to be performed separately by each thread – or one after the other – by using the command *critical* (or *atomic* in some cases). The user should try and reduce these areas in code as much as possible. For example, if the maximum of a given area is to be found, then it would be better if each thread found the maximum of the sub-array that is relevant to the thread and then find the global maximum, rather than having each thread check after every iteration whether the current value is larger than the global maximum. For simple operations, OpenMP provides the user with a simple definition for these actions, using the command *reduction*.

## 2  Scientific Software Parallelization using OpenMP

### 2.1  Loop Parallelization and Scheduling using OpenMP in Scientific Codes

A large area of code portions that can be parallelized are loops, where there is no dependency between the commands used in each of the iterations. Due to the high frequency of these cases, OpenMP has dedicated a number of commands to handle loop parallelization, including scheduling commands that handle the different calls in loops to other cores. The two main scheduling methods in OpenMP are the dynamic scheduling method and the static scheduling method. These methods are different from each other in the way the calls within the loop are assigned to the different threads, and they represent a specific and important case to the way the user may utilize the OpenMP library in order to determine the relationship between the different threads.

Provided that the selected scheduling method is a static method, the call assignment among the different threads is performed, independently of the call's running time. In comparison, when using a dynamic scheduling method, the master thread is managing the loop parallelization at runtime, and each one of the slave threads performs the loop for a number of calls. When one of the threads finishes processing its allocated number of calls, it receives additional calls from the master thread that it needs to perform. As explained above, in this method, the running time is divided relatively equally between the different threads.

For each one of these options, the developer may define another parameter – *chunk* – which fixes the number of calls that OpenMP parallelization considers as one unit. In static scheduling, the thread calls will be assigned in the following way: The main thread will receive the first address segment in the loop calls, the second thread will receive the second address segment, the N thread will receive the Nth address segment, and then the first (master) thread will receive the N+1 address segment, until all loop calls are assigned (Round-Robin fashion). In most cases, this option is not used, since it is less flexible compared to dynamic scheduling, and it is not as simple compared to a static scheduling method that does not use address segments. There are some applications where using static scheduling while controlling the address segment size provides the maximal efficiency, but these are rare cases and we estimate they are not represented in scientific software around the world. If a dynamic scheduling method is used, the loop calls will be divided to a fixed size of the address segments (the *Chuck* value), and each time a thread finishes its task, it will be given another portion from the master thread. Using the *chunk* value, the developer can decide on the rate between the granularity and the load balancing.

In terms of load balancing, it is better to use the dynamic scheduling method with an address segment as small as possible. However, in the cases where the size of information chunks computed is bigger than the cache memory (L3), there is a significant advantage in using the static scheduling method, since this way, each processor receives only a portion of the memory and there are no double memory loads. Of course, this is only true where the nodes share RAM memory and do not share L3 cache memory.

As for the size of the *cache line*, it is obvious that reducing cache miss to a minimum will be performed by adjusting the size of the address segment used by

OpenMP to the *cache line* size used by the given processor. Thus, considering that OpenMP platform parallelizes mostly the computation loops of arrays, and considering that the arrays that are used in scientific software around the world use the data type REAL(8) – where each cell occupies 8 bytes of memory – the address segment will be equal to the *cache line* size divided by 8 bytes (for example, in most cases of supercomputers in the world, the size of the *cache line* is 64 bytes, which means the optimal address segment size should be 8, assuming there is no vectorization). If parallelization is performed on the innermost part of the array (i.e., parallelizing the columns in an array using FORTRAN language), the problem of false-sharing is solved trivially using these *chunk* size. It is possible that while changing the order or arrays and loops will make the parallelization more efficient and reduce the total running time. It should be stated that in the static scheduling method the size of the *cache line* is directly affecting the efficiency of usage with different address segments sizes (provided that the user has decided to define such segments).

## 2.2 Computation Parallelization Commands

The number of parallelization commands in OpenMP is significantly smaller compared to the number of parallelization commands in MPI library, while some specific parallelization commands take up considerable portions of scientific software parallelization in the world. These commands – despite their simplicity and sometimes because of it – depict numerous risks to the validity of computations, due to the usage of a shared memory model.

However, these risks can be eliminated by following a few guidelines: During the process of parallelization in OpenMP, there is a need to make a strict separation between private variables, which each thread will create a copy of them, and shared variables, which will be shared between the different threads. Generally, variables whose value changes within a parallelized code segment need to be private variables, unless it is certain that only a single thread will change their value, or that their value change will be performed in a synchronized and correct manner. In *Do* loop parallelization, variables whose values are explicitly set within the loop must be private variables (since all threads will access them simultaneously). In comparison, an array may be a shared variable if each thread changes a different element in the array. The variables on which the shared action between different loops is being performed require performing synchronization actions between the loops at some point during the loop. As mentioned above, OpenMP allows performing such synchronization using the command *reduction*.

In the example below, we see standard loop parallelization in LEEOR2D with usage of *reduction*. The loop iterates a two-dimensional array (lines 2.1-2.1.1), when the parallelization is performed on index j (in other words, each thread goes over different values of *j*). Since every thread goes over *i* values independently, *i* index needs to be a private variable that is not shared between the threads (which is the way it is defined in the first line). Inside the loop, first the value of x is being set, based on *f*, a function of the double indices (line 2.1.1.1). Since this value is declared explicitly, it must be a private variable. Then, the index *i,j* changes in array *a* (line 2.1.1.2). Since each thread changes different portions of *a*, array *a* needs to be a shared variable. The variable *sum* accumulates all values of array *a* (line 2.1.1.3). Since the purpose of this code is to sum all values in array *a*, and not just the portion of *a* being handled by each thread separately, the *sum* variable needs to be defined as a variable that needs to go through *reduction*. This definition means that each one of the threads will create a private copy of the variable and accumulate the values it has iterated. When the loop terminates, the threads will accumulate together the private variables they each accumulated in order to obtain the total sum. This instruction is compiled using the expression +*:sum*.

Sum all values in array *a* using OpenMP.

```
1.  !$omp parallel do private(i,x) &
2.  !$omp reduction(+:sum)
    2.1.      do j = 1,NY
       2.1.1.      do i = 1,NX
          2.1.1.1.       x = f(i,j)
          2.1.1.2.       a(i,j) = g(x)
          2.1.1.3.       sum = sum + a(i,j)
       2.1.2.      end do
    2.2.      end do
3.  !$omp end parallel do
```

The next example demonstrates a common situation in LEEOR2D and other scientific software around the world where the usage of *reduction* is not sufficient, and there is a need to use *critical*. In this case, we look for the maximum value of array *a*, and the index where this value is obtained. While finding the maximum value may be done using *reduction,* reduction cannot be used for the index that obtains the value. In line 3, it is set that the initial value of the variable *priv_max* is very small. The parallelized loop is between lines 4 and 5, and after its termination, every thread knows the maximum value for the segment iterates, and which index obtains this value. Then, we need to find the maximum value between the different threads. This action is performed using *critical* to make sure that each thread updates the value of *max* separately, and to prevent a situation where a thread declares that its own maximum is the total maximum when it may not be the case.

Get the maximum value of array *a*, and the index where this value is obtained using OpenMP.

```
1.  max = -1d20
2.  !$omp parallel private(priv_max, priv_maxi)
3.  priv_max = -1d20
4.  !$omp do
    4.1.      do i = 1,NX
       4.1.1.      if(priv_max.lt.a(i)) then
          4.1.1.1.       priv_max  = a(i)
          4.1.1.2.       priv_maxi = i
       4.1.2.      end if
    4.2.      end do
5.  !$omp end do
```

```
6. !$omp critical
   6.1.     if(max.lt.priv_max) then
      6.1.1.       max  = priv_max
      6.1.2.       maxi = priv_maxi
   6.2.     end if
7. !$omp end critical
8. !$omp end parallel
```

## 3  AutOMP: A Pre-processor Parallelization Commands Software

While parallelizing the *do* loops in LEEOR2D and other scientific software around the world, it was found that except for specific factors whose distributions in the code is negligible and that the user should identify and handle separately (for example, the abovementioned *critical* code segments), most of the parallelization task can be performed using an ad-hoc pre-processor software. The role of the pre-processor software will be to identify common code patterns, process them and finally generate the precise parallelization command necessary for the user in order to parallelize the desired *do* loop.

In order to write pre-processor software that can generate OpenMP commands for scientific software (AutOMP.py), Python programming language is used; the software uses two variables in the command line: The first is a path to the file that contains the *do* loop, and the second path is to the file that contains a list of all of the variables in the module that contains the *do* loop. Below, we demonstrate the capabilities of this software, and the way it been used by executing it on a main *do* loop method from LEEOR2D.

Using AutOMP with the two input file paths.

```
AutOMP.py subroutine.f90 private_args.f90
```

First, the software filters the lines of code and categorizes - according to a variable definition file (private_args.f90) - the list of private variables of the subroutine in alphabetical order. This identification of private variables is performed in order to allow the OpenMP library to understand which variables need to be preserved with a distinct value for each one of the threads. The rest of the variables are defined as shared with the other threads by default.

Output No. 1 of the AutOMP operation.

```
All Private Arguments in the Declaration:
['A', 'AB', 'AL', 'AR', 'AR1', 'AR2', 'AT1', 'AT2', 'B',
'COFEL', 'DC_BMM' ... 'YCT', 'YP1', 'YP2', 'YP3', 'YP4',
'YSTMM', 'YY1', 'YY2', 'Z1', 'Z2', 'Z3', ... , 'Z9']
```

This list is not sufficient for the OpenMP library since it contains **all** variables of the subroutine, and not just the *do* loop variables that we would like to parallelize. Hence, after the first step, the software searches for all expressions of variable assignments with the *do* loop, whether it is a direct assignment ("YP4=0"), an assignment inside a *do* sub-loop, ("Do II=1, NMATS"), or by sending variables to an external subroutine ("CALL VOLMAT(XFLAG, FT(i)"). From these expressions, the software categorizes these variables into a separate list. This list essentially contains all variables in which the *do* loop uses, including the shared variables.

Output No. 2 of the AutOMP operation.

```
All Arguments in the Scope:
['AB', 'AL', 'AR', 'AR1', 'AR2', 'AT', 'AT1', 'AT2',
'BLABLA', 'DC_BMM' ... 'XMM2', 'XP1', 'XP2', 'XP3', 'XP4',
'Y1', 'Y2', 'Y3', 'Y4', 'YCR', 'YCT', 'YP1', 'YP2', 'YP3',
'YP4', 'YSTMM']
```

The program assumes that the parallelization is performed by dividing the data network to the number of given threads by sharing **one (and not more)** of the index variables in the network (I or J), and thus the variables that depend on the two index variables of the network are declared immediately after as shared variables and are not included in the private variables list.

Given the list of all private variables in the subroutine and all variables defined in the *do* loop, the intersection of these lists will provide the list of all variables that can be defined as private in the *do* loop.

Output No. 3 of the AutOMP operation.

```
Intersection Between Declaration and Scope Arguments:
['AB', 'AL', 'AR', 'AR1', 'AR2', 'AT1', 'AT2', 'BLABLA',
'DC_BMM', 'FB', 'FBM' ... 'XMM2', 'XP1', 'XP2', 'XP3',
'XP4', 'Y1', 'Y2', 'Y3', 'Y4', 'YCR', 'YCT', 'YP1', 'YP2',
'YP3', 'YP4', 'YSTMM']
```

However, this list also includes variables that are not shared or private, but variables of subtraction, multiplication or addition (*reduction*), and therefore the threads need to synchronize between each other on the values of these variables prior to any loop, since these values are meaningful for all threads ("VOLPH=VOLPH+VOLM(II)"). These values are omitted from the private variables list, and they are defined, along with their operator, separately.

Output No. 4 of the AutOMP operation.

```
Reduction Clauses Args:
['VOLPH', 'TOTM', 'TOTSIE', 'XMR', 'XMT', 'XML', 'XMB',
'XM0',    'Z1'    ...   'Z9',    'XMVF_HAN',    'XMVFP_HAN',
'XMVF_HANDMP', 'XMVFP_HANDMP']
Reduction Clauses Operators:
['+', '-', '*', '+', '+', '+', '-', '-', '+', '+', '+',
'+', '+', '-', '+', '+', '*', '+', '+', '*', '+']
```

After this sequence of operations, the program generates the precise parallelization code lines according to OpenMP that the user needs to place before the *do* loop in order to parallelize it. The integrity of the program was tested successfully on a collection of 20 loops. As previously said, in some cases some adjustments may be necessary, according to the complexity of the loop and the necessary parallelization sophistication, but there is no other option that a memory leak will occur.

Output No. 5 of the AutOMP operation – The Automatic OpenMP Parallelization command.

```
Ready to OpenMP Parallel Code:
!$omp parallel do private &
(AB, AL, AR, AR1, AR2, AT1, AT2, ... Y3, Y4, YCR, YCT, YP1,
YP2, YP3, YP4, YSTMM) &
schedule(dynamic, 8)
!$omp reduction(+:VOLPH)
!$omp reduction(-:TOTM)
!$omp reduction(*:TOTSIE)...
```

## 4  Performance Tests

As previously stated, the present work was done with the LEEOR2D code, developed to solve the Euler conservation system in a hydrodynamic continuum, and designated to run on a Xeon Phi coprocessors. Using AutOMP we successfully managed to parallel all of the main methods of the code, and currently achieved a stable parallelization.

The current parallel LEEOR2D ran on a 32 cores computer with a total of 96Gb of memory and resulted x22.5 times faster than the serial code. The results were the same (about less than 1% deviation) over 5 different tests. No memory leaks were found, and there was not even one case in which the AutOMP falsely handled the variables parallel organization.

## 5  Summary

Scientific computing is a field that exists since the early 1960's. In this field, computers are being used to analyze complex scientific problems. A major development in this field occurred in the 1990's when parallelization was introduced. Today, parallelized codes take a major part in all scientific fields including physics, chemistry and biology. In many cases the computer is used to solve a partial differential equation (PDE) depend on time and space, both are being discretized, an example for these problems is solving the Euler equations in fluid dynamics.

OpenMP is a set of compiler directives and callable runtime library routines that extend Fortran (and separately, C and C++) to express shared-memory parallelism. OpenMP is supported on most platforms, processor architectures and operating systems. The relative ease of implementing OpenMP in existing codes makes it favorable choice for parallelization of legacy codes.

OpenMP is an implementation of multithreading parallelism paradigm. In this method, a master thread runs the program manually. At a given point, specified by *!$omp parallel* command, the master thread forks number of slave threads, disturbing the work among them, according to OpenMP directives. OpenMP allows users to determine the way the work will be distributed among the threads. Different methods of work sharing allow the users to optimize the parallelization, by getting a good load-balancing and a fine granularity.

A large area of code portions that can be parallelized by OpenMP is loops, where each iteration does not depend on previous iterations. This is the case in most of the loops over the space in many different programs. OpenMP has a set of directives devoted to loop parallelization. When parallelizing a loop each thread will carry out certain amount of the loop iterations. OpenMP gives the user ways to choose how the loop iterations will be divided among the threads, and this is done by the *schedule* clause. The two main scheduling methods in OpenMP are the dynamic scheduling method and the static scheduling method. In the dynamic method the master thread manages the work sharing at runtime, it sends each of the slaves threads number of calls, and each time a slave thread finishes its iterations he contacts the master thread to get the next work chunk. This method achieves good load-balancing, but repeated communication between threads may damage the overall performance. In the static method the iterations are divided independently of the runtime of each loop iteration. This method has much less communication between the treads, but the load-balancing may be affected, especially if some loop iterations are more complex than other loop iterations. In both scheduling methods OpenMP has another parameter – *chunk*. This parameter determines the minimal loop iterations to be allocated to a thread in each allocation.

Two other important clauses types that need to be used in loop parallelization is data sharing attribute clauses and synchronization clauses. Each variable in the parallelized loop can be private, meaning each of the threads will create its own copy of the variable, and only this thread will be able to update its private copy of the variable. The other option is a shared variable, where only one copy of it will be saved in the shared memory, and each of the threads could access and change it. When updating a variable in shared memory, the user must avoid synchronization errors. OpenMP gives the user the option to control the synchronization with some synchronization clause such as *barrier, atomic* and *critical*. An additional OpenMP variable attribute is the *reduction* attribute: If a variable is a reduction variable, each of the threads will make a private copy of the variable, and after the parallel scope is executed a global variable will be update by preforming the declared operation on the private values.

During the parallelization process of a loop the programmer should choose for each variable its data sharing attribute, and either or not it should be reduced. An automatic tool for doing so had been introduced in this paper. The tool is called AutOMP and is written in the Python programming language as a pre-processing software.

The default data sharing attribute of variables is shared. First, AutOMP identifies the variables that are being updated inside the loop, either by a direct assignment, an assignment by a sub-loop, or by an external subroutine. These variables could not be shared among the threads without declaring the way it should be synchronized, unless the variable that is being updated is an array element which depends on the loop index. After Identifying all the updated variables within the loop, AutOMP identify which one of them should be a private variable, meaning that its value is not immanent outside of the parallel scope, and which should be synchronized between the threads in order to get a certain value. For those variables AutOMP recognizes the operator which need to be used to reduce to the final value. At the end, AutOMP gives the full OpenMP command to be used for parallelizing the loop.

## Acknowledgments

This work was supported by the Lynn and William Frankel Center for Computer Science.